# Relativistic Correction of the Field Emission Current in the Fowler-Nordheim Formalism

S.O. Lebedynskyi[1,*], O.O. Pasko[2], R.I. Kholodov[1]

[1] *Institute of Applied Physics, National Academy of Sciences of Ukraine, 58, Petropavlivska St., 40000 Sumy, Ukraine*
[2] *Sumy State University, 2, Rimsky Korsakov St., 40007 Sumy, Ukraine*

As the title implies the article describes the possibility of taking into account the relativistic correction to the field current density of the field emission of electrons from the metal. The article provides the reader with some analytic generalization of the Fowler-Nordheim equation with the relativistic correction. The relativistic correction to the Fowler–Nordheim equation makes it possible to take into account the influence of the relativism on the field emission current. It is especially noted that the consideration of this correction is necessary in the case of sufficiently strong electric fields and relatively large interelectrode distances. It should be stressed that this correction is valid for fixed interelectrode distances that decrease with increasing electric field strength. It means that for the electric field strength of 0.1 to 1 Gv/m the interelectrode distance should not exceed values of 1 to 0.1 cm. First in the article it is spoken in detail about finding of the electron wave function. Next the field emission curent calculations are given. As a result the transmission coefficient of the potential step from the Klein-Gordon equation within the framework of the Fowler-Nordheim approximation is found. It is shown that in the case of the interelectrode distanse less than 1 cm, an analytical expression for the field electron emission current density is obtained. The conclusion that usually relativistic correction does not exceed a tenth of a percent is made. But in the case of the field electron emission from pulsars (where the work function and electric field strength are much higher) the contribution of the relativistic correction about 10 % has been established.

## 1. INTRODUCTION

The theory of electron emission in intense external electric fields was written by Fowler and Nordheim in the 20-ies of the last century [1]. Despite this, it is widely used in various branches of physics in our days too. Field emission currents not only from metals but also from astrophysical objects and nanomaterials are described by this theory [2-3]. Also this theory is widely used for describing the dark currents from the construction materials of the accelerating structures [4, 5, 10]. The authors of the article [1] mark that the emission becomes sensible for fields of rather more than $10^8$ V/m and already is very large for fields of the order of $10^9$ V/m. Individual case of the big interelectrode gaps should be noted. For the gap of about 1 cm and the electric field strength near $E = 10^8$ V/m the electron motion becomes relativistic: $\gamma = \frac{\varepsilon}{mc^2} \approx 2$. This means the need to use relativistic positions on such scales. That's why we will find the transmission coefficient of the potential step from the Klein-Gordon equation within the framework of the Fowler-Nordheim approximation.

The aim of this article is to generalize the Fowler-Nordheim theory to a relativistic case.

## 2. DERIVATION OF THE ELECTRON WAVE FUNCTION

The Klein-Gordon equation in the case of an external electric field can be written as:

$$\left[\frac{\partial^2}{\partial x^2} + \frac{(\varepsilon - V + eEx)^2}{\hbar^2 c^2} - \frac{m^2 c^2}{\hbar^2}\right]\psi = 0,$$

where $\varepsilon$ is the electron energy, $V$ is the barrier height, $-e$ is the electron charge, $E$ is the electric field strength, $\hbar$ is the Planck constant, $c$ is the speed of light, $m$ is the electron mass.

Let's write this equation in dimensionless coordinates, considering that

$$k^2 = \frac{2m}{\hbar^2}, \quad x_0 = \frac{V - \varepsilon_k}{eE}, \quad \varepsilon = mc^2 + \varepsilon_k,$$

$$y = \left(\frac{2meE}{\hbar^2}\right)^{\frac{1}{3}}(x - x_0), \quad R = \frac{1}{c^2}\left(\frac{eE\hbar}{4m^2}\right)^{\frac{2}{3}},$$

$$\left[\frac{\partial^2}{\partial y^2} + y + Ry^2\right]\psi = 0. \quad (1)$$

The Schrödinger equation (which was used by Fowler and Nordheim [1]) in dimensionless coordinates is written as

$$\left[\frac{\partial^2}{\partial y^2} + y\right]\psi = 0. \quad (2)$$

We will use the Fowler-Nordheim approach, so we need to satisfy the condition $Ry \ll 1$. This means that for electric field strength $E = 10^8 \div 10^9$ V/m [1] the interelectrode distance should not exceed $1 \div 0.1$ cm.

Then, the solution of equation (2) can be represented as $\psi = \psi_0 + \psi_1$, where $\Psi_0$ is the solution of the Fowler-Nordheim equation [1]. Equation (1) can be written as

$$\frac{\partial^2 \psi_0}{\partial y^2} + y\psi_0 + \frac{\partial^2 \psi_1}{\partial y^2} + y\psi_1 + Ry^2(\psi_0 + \psi_1) = 0. \quad (3)$$

---
[*] lebos@nas.gov.ua

According to the equation (2)

$$\frac{\partial^2 \psi_0}{\partial y^2} + y\psi_0 \equiv 0.$$

At the same time, from [1] we know that
$\psi_0 = \sqrt{y} H^{(2)}_{\frac{1}{3}}\left(\frac{2}{3} y^{\frac{3}{2}}\right)$. Therefore, given the fact that $\Psi_0 \gg \Psi_1$, equation (3) can be rewritten in the form:

$$\frac{\partial^2 \psi_1}{\partial y^2} + y\psi_1 = f(y), \quad (4)$$

where $f(y) = -Ry^2 \Psi_0$.

To solve the equation (4), we will use the method of variation of parameters. Consider the solution of the equation $\frac{\partial^2 \psi'}{\partial y^2} + y\psi' = 0$. We already know [1] that the solution of this equation is the following:

$$\psi' = \sqrt{y} H^{(2)}_{\frac{1}{3}}\left(\frac{2}{3} y^{\frac{3}{2}}\right).$$

Then the solution of equation (4) can be written as

$$\psi_1 = C_1(y) g_1(y) + C_2(y) g_2(y), \quad (5)$$

where

$$g_1(y) = \sqrt{y} H^{(2)}_{\frac{1}{3}}\left(\frac{2}{3} y^{\frac{3}{2}}\right), \quad g_2(y) = \sqrt{y} H^{(2)}_{\frac{1}{3}}\left(\frac{2}{3} i y^{\frac{3}{2}}\right)$$

are the solutions of the homogeneous differential equation.

To find the $C_1(y)$ and $C_1(y)$, it is necessary to solve the system

$$\begin{cases} C_1'(y) g_1(y) + C_2'(y) g_2(y) = 0, \\ C_1'(y) g_1(y)' + C_2'(y) g_2(y)' = f(y). \end{cases}$$

The wave function must represent a wave that moves to the right (to converge with large y). We can find that $C_2(y) g_2(y)$ does not satisfy the condition if we plot graphs of functions $C_1(y) g_1(y)$, $C_2(y) g_2(y)$. Therefore, the solution of equation (4) is

$$\psi_1 = \left(-\int \frac{g_2 f(y)}{g_1(y) g_2'(y) - g_2(y) g_1'(y)} dy\right) \cdot \sqrt{y} H^{(2)}_{\frac{1}{3}}\left(\frac{2}{3} y^{\frac{3}{2}}\right). \quad (6)$$

Let's find an explicit expression of $C_1(y)$. To do this, let's write $\psi_1$ considering (5)

$$C_1(y) = \int \frac{Ry^2 dy}{\alpha_1 + i\beta_1 - \alpha_2 - i\beta_2},$$

where

$$\alpha_1 + i\beta_1 = \frac{\left[H^{(2)}_{\frac{1}{3}}\left(\frac{2}{3} i y^{\frac{3}{2}}\right)\right]'}{H^{(2)}_{\frac{1}{3}}\left(\frac{2}{3} i y^{\frac{3}{2}}\right)}, \quad \alpha_2 + i\beta_2 = \frac{\left[H^{(2)}_{\frac{1}{3}}\left(\frac{2}{3} y^{\frac{3}{2}}\right)\right]'}{H^{(2)}_{\frac{1}{3}}\left(\frac{2}{3} y^{\frac{3}{2}}\right)}$$

The Hankel function can be represented in the following form by definition [6]:

$$H^{(2)}_{\frac{1}{3}}(z) = \frac{i}{\sin\frac{\pi}{3}} \left\{ J_{-\frac{1}{3}}(z) - e^{\frac{\pi i}{3}} J_{\frac{1}{3}}(z) \right\};$$

$$H^{(2)}_{\frac{1}{3}}(iz) = \frac{-1}{\sin\frac{\pi}{3}} \left\{ I_{-\frac{1}{3}}(z) - e^{\frac{\pi i}{3}} I_{\frac{1}{3}}(z) \right\}.$$

It is possible to calculate $\alpha_1$ and $\beta_1$

$$\alpha_1 + i\beta_1 = \sqrt{y} \frac{I'_{-\frac{1}{3}}\left(\frac{2}{3} y^{\frac{3}{2}}\right) I_{-\frac{1}{3}}\left(\frac{2}{3} y^{\frac{3}{2}}\right) + \frac{1}{2} I'_{-\frac{1}{3}}\left(\frac{2}{3} y^{\frac{3}{2}}\right) I_{\frac{1}{3}}\left(\frac{2}{3} y^{\frac{3}{2}}\right)}{\left|-\sin\frac{\pi}{3} H^{(2)}_{\frac{1}{3}}\left(\frac{2}{3} y^{\frac{3}{2}}\right)\right|^2} +$$

$$+\sqrt{y} \frac{\frac{1}{2} I'_{\frac{1}{3}}\left(\frac{2}{3} y^{\frac{3}{2}}\right) I_{-\frac{1}{3}}\left(\frac{2}{3} y^{\frac{3}{2}}\right) + I'_{\frac{1}{3}}\left(\frac{2}{3} y^{\frac{3}{2}}\right) I_{\frac{1}{3}}\left(\frac{2}{3} y^{\frac{3}{2}}\right)}{\left|-\sin\frac{\pi}{3} H^{(2)}_{\frac{1}{3}}\left(\frac{2}{3} y^{\frac{3}{2}}\right)\right|^2} +$$

$$+i\sqrt{y} \frac{\frac{\sqrt{3}}{2}\left(I'_{\frac{1}{3}}\left(\frac{2}{3} y^{\frac{3}{2}}\right) I_{-\frac{1}{3}}\left(\frac{2}{3} y^{\frac{3}{2}}\right) - I'_{-\frac{1}{3}}\left(\frac{2}{3} y^{\frac{3}{2}}\right) I_{\frac{1}{3}}\left(\frac{2}{3} y^{\frac{3}{2}}\right)\right)}{\left|\sin\frac{\pi}{3} H^{(2)}_{\frac{1}{3}}\left(\frac{2}{3} y^{\frac{3}{2}}\right)\right|^2}$$

For evaluating $\alpha_1$ we should use the corresponding asymptotic values. It's easily to see that $\alpha_1 = \sqrt{y}$. Considering $\beta_1$ we can use the expression for the the Wronskian of Bessel's functions of purely imaginary argument:

$$W\left\{I_{\frac{1}{3}}, I_{-\frac{1}{3}}\right\} = -\frac{2\sin\frac{\pi}{3}}{\pi \frac{2}{3} y^{\frac{3}{2}}} = -\frac{\sqrt{3}}{\pi \frac{2}{3} y^{\frac{3}{2}}}.$$

Further we will use the asymptotic expression [6] (when Q is large) for denominator:

$$H^{(2)}_{\frac{1}{3}}(z) \sim \sqrt{\frac{2}{\pi z}} e^{-i\left(z - \frac{5\pi}{12}\right)}, \quad (7)$$

Then,

$$\beta_1 = -\sqrt{y} e^{-\frac{4}{3} y^{\frac{3}{2}}}.$$

It remains to calculate $\alpha_2$ and $\beta_2$:

$$\alpha_2 + i\beta_2 = \sqrt{y}\frac{J'_{-\frac{1}{3}}\left(\frac{2}{3}y^{\frac{3}{2}}\right)J_{-\frac{1}{3}}\left(\frac{2}{3}y^{\frac{3}{2}}\right) - \frac{1}{2}J'_{-\frac{1}{3}}\left(\frac{2}{3}y^{\frac{3}{2}}\right)J_{\frac{1}{3}}\left(\frac{2}{3}y^{\frac{3}{2}}\right) - \frac{1}{2}J'_{\frac{1}{3}}\left(\frac{2}{3}y^{\frac{3}{2}}\right)J_{-\frac{1}{3}}\left(\frac{2}{3}y^{\frac{3}{2}}\right) + J'_{\frac{1}{3}}\left(\frac{2}{3}y^{\frac{3}{2}}\right)J_{\frac{1}{3}}\left(\frac{2}{3}y^{\frac{3}{2}}\right)}{\left|-i\sin\frac{\pi}{3}H^{(2)}_{\frac{1}{3}}\left(\frac{2}{3}y^{\frac{3}{2}}\right)\right|^2} +$$

$$+i\sqrt{y}\frac{\frac{\sqrt{3}}{2}\left(J'_{-\frac{1}{3}}\left(\frac{2}{3}y^{\frac{3}{2}}\right)J_{\frac{1}{3}}\left(\frac{2}{3}y^{\frac{3}{2}}\right) - J'_{\frac{1}{3}}\left(\frac{2}{3}y^{\frac{3}{2}}\right)J_{-\frac{1}{3}}\left(\frac{2}{3}y^{\frac{3}{2}}\right)\right)}{\left|\sin\frac{\pi}{3}H^{(2)}_{\frac{1}{3}}\left(\frac{2}{3}y^{\frac{3}{2}}\right)\right|^2}$$

(8)

We will use the similar methods for evaluating $\alpha_2$ and $\beta_2$. Therefore, we can write that $\alpha_2 = 0$. And the expression for the Wronskian of Bessel's functions is:

$$W\left\{J_{\frac{1}{3}}, J_{-\frac{1}{3}}\right\} = -\frac{2\sin\frac{\pi}{3}}{\pi z} = -\frac{\sqrt{3}}{\pi z}. \quad (9)$$

Substituting the asymptotic expression (7) and the Wronskian (9) into equation (8), we can find that $\beta_2 = -\sqrt{y}$.

Now we can write the expression for $C_1(y)$ with sufficient precision:

$$C_1(y) = R\int\frac{y^{\frac{3}{2}}dy}{1+i}.$$

We can substitute this expression after integration into equation (6) and obtain an explicit form of the wave function $\Psi_1$

$$\psi_1 = \frac{Ry^3}{5}(1-i)H^{(2)}_{\frac{1}{3}}\left(\frac{2}{3}y^{\frac{3}{2}}\right).$$

And now we can write the wave function of an electron in vacuum keeping in mind that $\Psi = \Psi_0 + \Psi_1$:

$$\psi = \sqrt{y}H^{(2)}_{\frac{1}{3}}\left(\frac{2}{3}y^{\frac{3}{2}}\right)\left[1 + \frac{Ry^{\frac{5}{2}}}{5}(1-i)\right].$$

The wave function of an electron inside metal is [1]:

$$\psi = \frac{1}{W^{\frac{1}{4}}}\left[ae^{-ikx\sqrt{W}} + a'e^{ikx\sqrt{W}}\right]. \quad (10)$$

### 3. CALCULATIONS OF THE FIELD EMISSION CURRENT

The same symbols as in article [1] will be used in the further calculations of $V = C$, $\varepsilon_k = W$, $eE = F$. Now we need to stitch two solutions.
The boundary conditions at $x = 0$ are:
1. $\Psi$ is always continuous;
2. $\partial\Psi/\partial x$ is continuous except the points where the potential is infinite.
At $x = +0$ we then find:

$$\psi(0) = \left(\frac{C-W}{F}\right)^{\frac{1}{2}}e^{-\frac{\pi i}{2}}\left(k^2F\right)^{\frac{1}{6}}H^{(2)}_{\frac{1}{3}}\left(\frac{2}{3}e^{-\frac{3\pi i}{2}}k\sqrt{F}\left(\frac{C-W}{F}\right)^{\frac{3}{2}}\right)\cdot\left[1 + e^{-\frac{5\pi i}{2}}\left(k^2F\right)^{\frac{5}{6}}\left(\frac{C-W}{F}\right)^{\frac{5}{2}}\frac{R(1+i)}{5}\right] \quad (11)$$

$$\left(\frac{\partial\psi}{\partial x}\right)_0 = \begin{bmatrix}\frac{1}{2}\left(\frac{C-W}{F}\right)^{-\frac{1}{2}}e^{\frac{\pi i}{2}}\left(k^2F\right)^{\frac{1}{6}}H^{(2)}_{\frac{1}{3}}\left(\frac{2}{3}e^{-\frac{3\pi i}{2}}k\sqrt{F}\left(\frac{C-W}{F}\right)^{\frac{3}{2}}\right) - \\ \left(\frac{C-W}{F}\right)^{\frac{1}{2}}\left(k^2F\right)^{\frac{1}{6}}k(C-W)H^{(2)'}_{\frac{1}{3}}\left(\frac{2}{3}e^{-\frac{3\pi i}{2}}k\sqrt{F}\left(\frac{C-W}{F}\right)^{\frac{3}{2}}\right)\end{bmatrix}\cdot$$

$$\cdot\left[1 + e^{-\frac{5\pi i}{2}}\left(k^2F\right)^{\frac{5}{6}}\left(\frac{C-W}{F}\right)^{\frac{5}{2}}\frac{R(1+i)}{5}\right] + \frac{R(1-i)}{2}\left(\frac{C-W}{F}\right)^2\left(k^2F\right)H^{(2)}_{\frac{1}{3}}\left(\frac{2}{3}e^{-\frac{3\pi i}{2}}k\sqrt{F}\left(\frac{C-W}{F}\right)^{\frac{3}{2}}\right)$$

We should equate these values to $\psi$ and $\partial\Psi/\partial x$ derived from (10) in $x = –0$.

The penetration coefficient of the potential step determines by the relation $\frac{|a|^2}{|a'|^2}$ and we can simplify the

resulting equation by leaving out factors which are present both in a and a':

$$\frac{2}{3}\left(k^2F\right)^{\frac{1}{2}}\left(\frac{C-W}{F}\right)^{\frac{3}{2}} = Q,$$

and Q is real and relatively large. In addition,

$$H^{(2)'}_{\frac{1}{3}}\left(e^{-\frac{3\pi i}{2}}Q\right) = e^{\frac{3\pi i}{2}} \frac{dH^{(2)}_{\frac{1}{3}}}{dQ}\left(e^{-\frac{3\pi i}{2}}Q\right).$$

$$a+a' = W^{\frac{1}{4}}\left(\frac{C-W}{F}\right)^{\frac{1}{2}} H^{(2)}_{\frac{1}{3}}\left(e^{-\frac{3\pi i}{2}}Q\right)\cdot\left[1 - \frac{R(1+i)}{5}\left(k^2 F\right)^{\frac{5}{6}}\left(\frac{C-W}{F}\right)^{\frac{5}{2}}\right] \quad (12)$$

$$-a+a' = \frac{i}{kW^{\frac{1}{4}}}\left[\frac{1}{2}\left(\frac{C-W}{F}\right)^{-\frac{1}{2}} H^{(2)}_{\frac{1}{3}}\left(e^{-\frac{3\pi i}{2}}Q\right) + \left(\frac{C-W}{F}\right)k\sqrt{F}\frac{dH^{(2)}_{\frac{1}{3}}}{dQ}\left(e^{-\frac{3\pi i}{2}}Q\right)\right]\cdot$$

(13)

$$\cdot\left[1 + e^{-\frac{\pi i}{2}}\frac{R(1-i)}{5}\left(k^2 F\right)^{\frac{5}{6}}\left(\frac{C-W}{F}\right)^{\frac{5}{2}}\right] + \frac{1}{kW^{\frac{1}{4}}}\frac{R(1-i)}{2}\left(\frac{C-W}{F}\right)^2\left(k^2 F\right)^{\frac{5}{6}} H^{(2)}_{\frac{1}{3}}(Q)$$

We will use the following relation to express the second-kind Hankel function $H^{(2)}_{\frac{1}{3}}\left(e^{-\frac{3\pi i}{2}}Q\right)$ in terms of real functions $I_{\pm\frac{1}{3}}(Q)$:

$$H^{(2)}_{\frac{1}{3}}\left(e^{-\frac{3\pi i}{2}}Q\right) = \frac{-1}{\sin\left(\frac{\pi}{3}\right)}\left(I_{-\frac{1}{3}}(Q) + e^{\frac{\pi i}{3}} I_{\frac{1}{3}}(Q)\right). \quad (14)$$

And now we can write

$$\alpha + i\beta = \frac{I'_{-\frac{1}{3}}(Q) + e^{\frac{\pi i}{3}} I'_{\frac{1}{3}}(Q)}{I_{-\frac{1}{3}}(Q) + e^{\frac{\pi i}{3}} I_{\frac{1}{3}}(Q)}, \quad (15)$$

where $\alpha$ and $\beta$ are real. Now we can write the transmission coefficient of the potential barrier under the influence of the electric field with relativistic corrections. We can use (14) and (15) for solving equations (12-13) and we will find

$$D = \frac{|a|^2 - |a'|^2}{|a|^2} =$$

$$= \frac{4\beta\left(\frac{C-W}{F}\right)^{\frac{3}{2}}\sqrt{F}\left(1 - \frac{2}{5}R\left(k^2 F\right)^{\frac{5}{6}}\left(\frac{C-W}{F}\right)^{\frac{5}{2}}\right)}{\left[\left\{W^{\frac{1}{4}}\left(\frac{C-W}{F}\right)^{\frac{1}{2}} + \frac{C-W}{W^{\frac{1}{4}}\sqrt{F}}\beta - \frac{R}{5kW^{\frac{1}{4}}}\left[\left(k^2 F\right)^{\frac{4}{3}}\left(\frac{C-W}{F}\right)^{\frac{7}{2}}(\alpha+\beta) + \frac{\left(k^2 F\right)^{\frac{5}{6}}\left(\frac{C-W}{F}\right)^2}{2} + \left(k^2 F\right)^{\frac{5}{6}}\left(\frac{C-W}{F}\right)^3 kW^{\frac{1}{2}}\right]\right\}^2 + \right.} \quad (16)$$

$$\left. + \frac{1}{k^2 W^{\frac{1}{2}}}\left\{\frac{1}{2}\left(\frac{C-W}{F}\right)^{-\frac{1}{2}} + \frac{C-W}{\sqrt{F}}k\alpha + \frac{R\left(k^2 F\right)^{\frac{5}{6}}\left(\frac{C-W}{F}\right)^2}{5}\left(\left(\frac{C-W}{F}\right)kW^{\frac{1}{2}} - 3\right) - \frac{R}{5}\left(k^2 F\right)^{\frac{4}{3}}\left(\frac{C-W}{F}\right)^{\frac{7}{2}}(\alpha-\beta)\right\}^2\right]$$

We can find that $\beta$ in (16) is

$$\beta = \frac{\sqrt{3}}{2}\frac{\left(I'_{\frac{1}{3}}(Q)I_{-\frac{1}{3}}(Q) - I'_{-\frac{1}{3}}(Q)I_{\frac{1}{3}}(Q)\right)}{\left|\sin\frac{\pi}{3} H^{(2)}_{\frac{1}{3}}\left(e^{-\frac{3\pi i}{2}}Q\right)\right|^2}.$$

Also we can see that the numerator is the Wronskian of the modified Bessel function of the first kind:

$$I'_{\frac{1}{3}}(Q)I_{-\frac{1}{3}}(Q) - I'_{-\frac{1}{3}}(Q)I_{\frac{1}{3}}(Q) = -\frac{2\sin\frac{\pi}{3}}{\pi Q}.$$

We will use the asymptotic expansion when Q is large for the denominator. Therefore,

$$\left|\sin\frac{\pi}{3} H^{(2)}_{\frac{1}{3}}\left(e^{-\frac{3\pi i}{2}}Q\right)\right|^2 \sim \frac{3}{4}\frac{2}{\pi Q}e^{2Q},$$

so that

$$\beta \sim e^{-2Q}.$$

Using the asymptotic values, we can also find that $a = 1$. It is easy to see that the terms that independent of $k$ are dominant in the denominator (16). And we can write with sufficient accuracy that:

$$D(W) = \frac{4 e^{-\frac{4}{3}(k^2 F)^{\frac{1}{2}} \left(\frac{C-W}{F}\right)^{\frac{3}{2}}} (W(C-W))^{\frac{1}{2}}}{C} \cdot \left(1 + \frac{\sqrt{2}}{5}\left(\frac{C-W}{mc^2}\right)^{\frac{5}{2}} \left(\frac{E_s}{E}\right)\right), \quad (17)$$

where $E_s = \frac{m_e^2 c^3}{e\hbar}$ is the Schwinger limit.

It is easy to see that the expression for the transmission coefficient consists of two parts. The first part is nothing other than the transmission coefficient obtained in [1] by Fowler and Nordheim. The second part of expression (17) is a relativistic correction to this expression. Therefore, the formula for the current density of field emission can be written in an analogous way to article [1]:

$$j = j_{FN}\left(1 + \frac{\sqrt{2}}{5}\left(\frac{C-W}{mc^2}\right)^{\frac{5}{2}}\left(\frac{E_s}{E}\right)\right),$$

where $j_{FN} = \frac{e^3 \mu^{\frac{1}{2}} E^2}{4\pi^2 \hbar (\chi + \mu) \chi^{\frac{1}{2}}} \cdot e^{-\frac{4}{3}\frac{\sqrt{2m}}{\hbar}\frac{\chi^{\frac{3}{2}}}{eE}}$ is the current obtained by Fowler and Nordheim [1], $E$ is the electric field strength; $\chi$ is the work function; $\mu$ is the thermodynamic partial potential of an electron.

Fowler and Nordheim in their work [1] emphasize that commonly measurable field emission current starts for $E$ values of about $10^8$ V/m. To estimate the relativistic correction, we use the average value of the work function of metals of about $\chi \approx 4$ eV. Therefore, the relativistic correction will be about 0.064 %. Since the field emission current density increases exponentially, this effect will be hardly visible experimentally. But we can consider the electron field emission current, for example, from pulsars, which is also described by the Fowler-Nordheim equation [1]. In this case, when magnetic field strength is $B = 10^8$ T, the electric field strength is up to $E \approx 10^{12}$ V/m [8] and work function is $\chi \approx 100$ keV [9]. The relativistic correction will be about 10 % and can make a significant contribution to the field emission current density in this case.

## 4. CONCLUSIONS

This paper considers the possibility of taking into account the relativistic correction to the current density of the field emission of electrons from the metal. It is shown that the consideration of this correction is necessary in the case of sufficiently strong electric fields and relatively large interelectrode distances. An analytic generalization of the Fowler-Nordheim equation with the relativistic correction was made. In the case of the interelectrode distance less than 1 cm, an analytical expression for the field electron emission current density is obtained. It is shown that usually relativistic correction does not exceed a tenth of a percent. But in the case of the field electron emission from pulsars (where the work function and electric field strength are much higher), the contribution of the relativistic correction is about 10 %.

### ACKNOWLEDGMENTS

Publication is based on the research provided by the grant support of the State Fund For Fundamental Research as well as by the National Academy of Sciences of Ukraine (NASU) under the program of cooperation between NASU, CERN and JINR «Fundamental research on high-energy physics and nuclear physics (international cooperation)» grant II – 32 –18.### REFERENCES

1. R. H. Fowler, L. Nordheim, *Proc. Roy. Soc. London A* **119**, 173 (1928).
2. A. Ghosh, S. Chakrabarty, *Mon. Not. R. Astron. Soc.*, **425**, (2012).
3. F. Giubileo, A. Di Bartolomeo, L. Iemmo, G. Luongo, F. Urban, *Appl. Sci.*, **8(4)**, 526, (2018)
4. M. Kidemo, Nucl. Instrum. *Methods Phys. Res. A* **530**, 596 (2004).
5. S. Lebedynskyi, O. Karpenko, R. Kholodov, V. Baturin, Ia. Profatilova, N. Shipman, W. Wuensch, *Nuclear Inst. and Methods in Physics Research*, A **908**, 318 (2018).
6. Crowder, M. (2011), NIST Handbook of Mathematical Functions (International Statistical Review: 2011)
7. V. S. Beskin, A. V. Gurevich, Ya. N. Istomin, *Physics of the pulsar magnetosphere* (Cambridge University Press:1993).
8. D. A. Diver, A. A. da Costa, E. W. Laing, C. R. Stark, L. F. A. Teodoro, *Mon. Not. R. Astron. Soc. 401*, 613 (2010).
9. N. Shipman, Ia. Profatilova, A.T. Perez Fontenla, W. Wuensch, S. Calatroni, Effect of an Externally Applied Magnetic Field on the Breakdown Rate in Ultra-High Vacuum Measured in the Large Electrode System at CERN, CLIC-Note-**1078**, 1 (2017).

## Релятивістська корекція струму польової емісії у формалізмі Фаулера-Нордгейма

С.О. Лебединський[1], О.О. Пасько[2], Р.І. Холодов[1]

[1] *Інститут прикладної фізики Національної академії наук України, вул. Петропавлівська, 58, 40000 Суми, Україна*
[2] *Сумський державний університет, вул. Римського-Корсакова, 2, 40007 Суми, Україна*

У статті описана можливість врахування релятивістської поправки до густини струму польової емісії електронів з металу. Представлено деяке аналітичне узагальнення рівняння Фаулера-Нордгейма з релятивістською поправкою. Релятивістська поправка рівняння Фаулера-Нордгейма дозволяє врахувати вплив релятивізму на польовий емісійний струм. Відзначимо, що врахування цієї поправки є необхідним у випадку досить сильних електричних полів і відносно великих міжелектродних відстаней. Слід підкреслити, що ця поправка справедлива для фіксованих міжелектродних відстаней, які зменшуються зі збільшенням напруженості електричного поля. Це означає, що для напруженості електричного поля від 0.1 до 1 Гв/м відстань між електродами не повинна перевищувати значень від 1 до 0.1 см. Спочатку в статті детально йдеться про знаходження хвильової функції електрону. Далі наводяться розрахунки польової емісії електронів. У результаті знайдено коефіцієнт проходження потенціального бар'єру з рівняння Клейна-Гордона в рамках наближення Фаулера-Нордгейма. Отримано аналітичний вираз для густини струму поля електронів у випадку міжелектродної відстані менше 1 см. Зроблено висновок, що зазвичай релятивістська поправка не перевищує десятої частини відсотка. Але у випадку польової електронної емісії з пульсарів (де робота виходу та напруженість електричного поля набагато вища) внесок релятивістської поправки становить близько 10%.

**Ключові слова:** Польова емісія, Темний струм, Релятивізм, Рівняння Клейна-Гордона, Рівняння Фаулера-Нордгейма.